\documentclass[preprint2, tighten]{aastex631}

\usepackage{amsmath}
\usepackage{amsfonts}

\received{XXX}
\revised{XXX}
\accepted{XXX}

\submitjournal{ApJ}

\shorttitle{Radio emitting outflows in IMBH hosting AGNs}
\shortauthors{Yang et al.}
\graphicspath{{./}{figures/}}

\begin{document}


\title{Radio observations of four active galactic nuclei hosting intermediate-mass black hole candidates: studying the outflow activity and evolution}

\correspondingauthor{Xiaolong Yang}
\email{yangxl@shao.ac.cn}

\author[0000-0002-4439-5580]{Xiaolong Yang}
\affiliation{Shanghai Astronomical Observatory, Key Laboratory of Radio Astronomy, Chinese Academy of Sciences, Shanghai 200030, China}
\affiliation{Shanghai Key Laboratory of Space Navigation and Positioning Techniques, Shanghai Astronomical Observatory, Chinese Academy of Sciences, Shanghai 200030, China}

\author[0000-0002-2211-0660]{Prashanth Mohan}
\affiliation{Shanghai Astronomical Observatory, Key Laboratory of Radio Astronomy, Chinese Academy of Sciences, Shanghai 200030, China}

\author[0000-0002-2322-5232]{Jun Yang}
\affiliation{Department of Space, Earth and Environment, Chalmers University of Technology, Onsala Space Observatory, SE-439\,92 Onsala, Sweden}

\author[0000-0001-6947-5846]{Luis C. Ho}
\affiliation{Kavli Institute for Astronomy and Astrophysics, Peking University, Beijing 100871, China}
\affiliation{Department of Astronomy, School of Physics, Peking University, Beijing 100871, China}

\author[0000-0002-0268-0375]{J.N.H.S. Aditya}
\affiliation{ARC centre for Excellence for All-Sky Astrophysics in 3 Dimensions (ASTRO 3D)}
\affiliation{Sydney Institute for Astronomy, School of Physics A28, The University of Sydney, NSW 2006, Australia}

\author[0000-0001-8485-2814]{Shaohua Zhang}
\affiliation{Shanghai Key Lab for Astrophysics, Shanghai Normal University, Shanghai 200234, People's Republic of China}

\author[0000-0002-5125-695X]{Sumit Jaiswal}
\affiliation{Shanghai Astronomical Observatory, Key Laboratory of Radio Astronomy, Chinese Academy of Sciences, Shanghai 200030, China}

\author[0000-0001-5323-0764]{Xiaofeng Yang}
\affiliation{Xinjiang Astronomical Observatory, Key Laboratory of Radio Astronomy, Chinese Academy of Sciences, 150 Science 1-Street, 830011 Urumqi, P.R. China}



\begin{abstract}
Observational searches for intermediate-mass black holes (IMBHs; $10^2 - 10^6$ $M_\odot$) include relatively isolated dwarf galaxies. For those that host active galactic nuclei (AGNs), the IMBH nature may be discerned through the accretion - jet activity. We present radio observations of four AGN-hosting dwarf galaxies (which potentially harbor IMBHs). Very large array (VLA) observations indicate steep spectra (indices of $-$0.63 to $-$1.05) between 1.4 and 9\,GHz. A comparison with the 9\,GHz in-band spectral index however shows a steepening for GH047 and GH158 (implying older/relic emission) and flattening for GH106 and GH163 (implying recent activity). Overlapping emission regions in the VLA 1.4\,GHz and our very long baseline array (VLBA) 1.5\,GHz observations, and possibly symmetric pc-scale extensions are consistent with recent activity in the latter two. Using the compact VLBA radio luminosity, X-ray luminosity (probing the accretion activity) and the black hole masses, all AGNs are found to lie on the empirical fundamental plane relation. The four AGN are radio quiet with relatively higher Eddington ratios ($0.04 - 0.32$) and resemble the X-ray binaries during spectral state transitions that entail an outflow ejection. Furthermore, the radio to X-ray luminosity ratio $\log{R_\mathrm{X}}$ of $-3.9$ to $-5.6$ in these four sources support the scenarios including corona mass ejection from accretion disk and wind activity. The growth to kpc-scales likely proceeds along a trajectory similar to young AGNs and peaked spectrum sources. The above complex clues can thus aid in the detection and monitoring of IMBHs in the nearby Universe.
\end{abstract}



\section{Introduction}\label{sec:intro}
Astrophysical black holes (BHs), inferred through their observational signatures are currently understood to be classified into two categories based on their mass. Stellar-mass BHs ($3 - 100$ $M_\odot$) originate from the end stages of the evolution of massive stars \citep{2017NewAR..78....1M}, as has been inferred from studies of X-ray binaries (XRBs; that can host a BH actively accreting from a companion star) in our Galaxy. Supermassive BHs (SMBHs; $\geq 10^6$ $M_\odot$) on the other hand are resident at the centers of most massive galaxies with bulges \citep{2013ARA&A..51..511K}. These have been mainly inferred through their role in the evolution of the host galaxy (through the correlations of the SMBH mass with the galactic bulge properties, including the dispersion velocity, luminosity, and mass). While there have been deductions of SMBH hosts even in the early Universe \citep[$\leqslant$ Gyr,][]{2015Natur.518..512W, 2018Natur.553..473B,2020NatCo..11..143A} through their observational signatures (accretion power and nuclear activity), modes of growth to such large masses ($10^6 - 10^{10}$ $M_\odot$) remain debatable \citep[see][]{2010A&ARv..18..279V,2020ARAA..58..257G}. Possibilities include mergers and accretion activity. If growing up from stellar-mass seed BHs, these scenarios would require extremely high accretion rates. On the other hand, the presence of intermediate-mass BHs (IMBHs; $10^2 - 10^6$ $M_\odot$) can help realize these scenarios more efficiently than the lower mass seed BHs, and can thus help in understanding SMBH formation and their influence on galaxy evolution \citep[see][and references therein]{2010A&ARv..18..279V,2017IJMPD..2630021M,2020ARAA..58..257G}.

The study of IMBHs is additionally important to related fields \cite[e.g.][]{2017IJMPD..2630021M,2020ARAA..58..257G}. These include understanding if IMBH systems follow a potential scale invariance of disk - jet activity \cite[accretion, jet/outflow ejection, i.e. spanning from stellar to supermassive, e.g.][]{1995A&A...293..665F,2003MNRAS.345.1057M, 2014ApJ...788L..22G}, their role in enabling and powering tidal disruption events \cite[e.g.][]{2016MNRAS.455..859S, 2018ApJ...867...20C}, including as wandering off-nuclear (in their host galaxy) sources \cite[e.g.][]{2019ApJ...871L...1T, 2021NatAs...5..560P,2021MNRAS.501.1413N,2020ApJ...888...36R}, and as contributors to the gravitational wave background when involving mergers \cite[e.g.][]{2021MNRAS.501.1413N}.

Based on the expected formation and evolution scenarios, observational searches for IMBHs have typically focused on the following habitats: globular clusters \citep[e.g.][]{2005ApJ...634.1093G, 2017Natur.542..203K, 2021ApJ...918...46W, 2022ApJ...924...48P}, ultra/hyper-luminous X-ray sources \citep[e.g.][]{2012Sci...337..554W, 2015ApJ...811L..11P}, and dwarf galaxies \citep[e.g.][]{2003ApJ...588L..13F, 2004ApJ...610..722G}. Among them, dwarf galaxies have recently received increasing attention owing to a number of them hosting AGNs \cite[e.g.][and references therein]{2022MNRAS.511.4109D}, allowing the identification of IMBHs based on their accretion signatures. These dwarf galaxies have undergone few merger events in their evolutionary history, and therefore, have not grown significantly since their birth \citep{2020ARAA..58..257G}. A large fraction of dwarf galaxies with masses $10^7-10^{10}\,M_\odot$ may then potentially host $10^4-10^6\,M_\odot$ black holes \citep[e.g.][]{2004ApJ...610..722G, 2007ApJ...670...92G, 2012ApJ...755..167D, 2013ApJ...775..116R, 2018ApJS..235...40L, 2018ApJ...863....1C, 2018MNRAS.478.2576M}.

Radio properties may be quantified in terms of the radio loudness parameter \cite[e.g.][]{1989AJ.....98.1195K} for comparison to optical properties and is given by $\mathcal{R}{\equiv}1.3\times10^5~L_\mathrm{5~GHz}/L_\mathrm{B}$, where $L_\mathrm{5GHz}$ and $L_\mathrm{B}$ are the 5\,GHz and optical $B$-band (4400\AA) monochromatic luminosities respectively. Synchrotron emission in radio-loud AGNs (RL-AGNs; $\mathcal{R} > 10$) ensues primarily from the acceleration of electrons to relativistic energies by physical processes involving a jet (collimated outflow) \cite[e.g.][]{1979ApJ...232...34B,2019ARA&A..57..467B}. The relativistic jet can span a wide range of physical scales (from less than a pc to a few ten - hundreds of kpc) and are ubiquitous in RL-AGNs, while a large portion of AGNs is radio-quiet (RQ, $\mathcal{R} < 10$). The absence of a prominent jet in RQ-AGNs allows probing of a wide range of physical mechanisms producing radio emission \cite[e.g.][and references therein]{2019NatAs...3..387P}. These can include sub-relativistic wide-angled winds \cite[e.g. ][]{2014MNRAS.442..784Z, 2015MNRAS.447.3612N,2021MNRAS.504.3823W}, lower power jets \cite[e.g.][]{2020MNRAS.494.1744Y,2021MNRAS.500.2620Y,2021MNRAS.504.3823W}, free-free emission from photo-ionized gas in the circum-nuclear region \citep{2021MNRAS.508..680B}, star-forming regions \cite[e.g.][]{2021MNRAS.500.2620Y}, and accretion disk - corona activity \cite[e.g.][]{2008MNRAS.390..847L, 2016MNRAS.459.2082R,2018ApJ...869..114I,2020ApJ...904..200Y}.

A pilot radio survey and study of low-mass AGNs with high accretion rates by \citet[][]{2006ApJ...636...56G} indicates that a large portion of these sources is predominantly radio-quiet. They draw a comparison to Galactic X-ray binaries (XRBs) where the high soft X-ray spectral state is characterized by a prominent accretion disk emission, and a quenched radio emission \cite[e.g.][]{2006csxs.book..157M}; for the AGN at high accretion rates, a similarity with XRBs \cite[if physical phenomena are scale independent, e.g.][]{2003MNRAS.343L..59H} may possibly explain their exceptionally low radio loudness. Indeed, the study of \cite{2020ApJ...904..200Y} finds that a large portion of AGNs accreting at high and super-Eddington rates (Eddington ratio $\lambda_{\rm Edd} = L_{\rm bol}/L_{\rm Edd} \gtrsim 1$, where $L_{\rm bol}$ is the bolometric luminosity and $L_{\rm Edd}$ is the Eddington luminosity) tend to be radio-quiet, confirming an inverse relationship between $\mathcal{R}$ and $\lambda_{\rm Edd}$ \citep{2002ApJ...564..120H}.

The origin of X-ray and radio emission from the vicinity of the compact central engine in XRBs (putatively from the accretion and jet components respectively) motivates an investigation of their dependence on the black hole mass ($M_{\rm BH}$) through a disk-jet coupling \cite[fundamental plane relation,][]{2003MNRAS.345.1057M}. The relation does not however clearly distinguish between the RL and RQ-AGN where the dominant emission mechanisms may differ; this may contribute in part to the scatter in the fitting. Further, the study of \citet{2008MNRAS.390..847L} finds that optically selected RQ-AGN from the Palomar-Green bright quasar survey \citep{1983ApJ...269..352S} indicates a constant $\log R_{\rm X} \approx -5$, independent of $M_{\rm BH}$, where $R_\mathrm{X}{\equiv}L_\mathrm{5~GHz}/L_\mathrm{X}$ ($L_\mathrm{X}$ is the 2 - 10 keV X-ray luminosity). The study of \citet{2014ApJ...788L..22G} however finds that a sample of low-mass AGNs ($\leqslant 10^{6.3}~M_\odot$) falls within the statistical scatter on the fundamental plane relation. The study of \cite{2022MNRAS.513.4673B} employs a statistically viable sample of RL and RQ-AGNs with radio and X-ray measurements to address this. They find that scatter in the fit (involving coefficients of $L_{\rm X}$, $M_{\rm BH}$ and normalization) for RQ-AGNs is consistent with the relation of \cite{2003MNRAS.345.1057M} but that for RL-AGNs largely diverges. The relation can then be used as a test for IMBHs hosted by low-mass RQ-AGNs, including in dwarf galaxies. Though, the studies of \citet{2006ApJ...636...56G}, \citet{2014ApJ...788L..22G} and \citet{2022MNRAS.513.4673B} employ arcsec-scale resolution Very Large Array (VLA) radio observations in investigating the properties of the low-mass and RQ-AGN and arriving at the above conclusions relating to the complex relations governing $\mathcal{R}$, $\lambda_{\rm Edd}$, and $M_{\rm BH}$.

Very Long Baseline Interferometry (VLBI) radio observations offer high angular resolutions (at the milli-arcsec scale), surpassing other imaging techniques in astronomy. The VLBI detection of compact pc-scale radio-emitting structures (core/core-jet/jet-knot) in the nuclear regions of dwarf galaxies can directly probe the jet/outflow activity enabled by an accreting, potential IMBH. To date, high-resolution VLBI observational studies are limited to only individual IMBH candidate hosts: NGC~4395 \citep[][]{2006ApJ...646L..95W}, Henize~2-10 \citep[][]{2012ApJ...750L..24R}, NGC~404 \citep[][]{2014ApJ...791....2P} and RGG~9 \citep[][]{2020MNRAS.495L..71Y}. Observations of NGC~4395 \cite[$R_{\rm X} = -5$, ][]{2003ApJ...583..145T} with the High Sensitivity Array (HSA) at 1.4 GHz reveal a radio quiet nucleus and an elongated sub-pc scale structure, indicative of an outflow \citep{2006ApJ...646L..95W}. The recent study of NGC~4395 by \citep{2022MNRAS.514.6215Y} finds an undetected source from European VLBI Network (EVN) 5 GHz high-resolution observations. Very large array (VLA) 15 GHz observations detect a core (coincident with the {\it Gaia} optical position) and eastern (E) components; a new HSA imaging confirms the E component that is interpreted as a propagating shock originating from episodic ejection or outflow activity. Observations of NGC\,404 with the EVN at 5 GHz \citep{2014ApJ...791....2P} results in a non-detection (sub-pc to pc scale), though extended structures over tens of pc have been detected by VLA 1.4 GHz observations \citep{2012ApJ...753..103N}; a contemporaneous {\it Chandra} X-ray observation indicates a non-variable source with $\log{R_\mathrm{X}}<-3.8$.

In this work, we present and discuss Very Long Baseline Array (VLBA) observational results from four IMBH candidates with masses $\sim10^5\,M_\odot$ (see Table \ref{tab:info}). Throughout the work, we adopt the standard $\Lambda$CDM cosmology with a Hubble constant $H_0 =70$ km s$^{-1}$ Mpc$^{-1}$, and matter density and dark energy density parameters $\Omega_\Lambda=0.73$, $\Omega_m=0.27$ respectively.

\section{The sample} \label{sec:sample}
The study of \citet{2004ApJ...610..722G} identified 19 IMBH candidates from the Sloan Digital Sky Survey (SDSS) Data Release 1 (DR1). The virial mass technique has been subsequently used to identify several hundred IMBH candidates \citep{2007ApJ...670...92G, 2012ApJ...755..167D, 2013ApJ...775..116R, 2018ApJS..235...40L} from the later SDSS data release. We compiled a list containing all targets (598 in total) from the above studies, then cross-matched it with the NRAO VLA Sky Survey \citep[NVSS,][]{1998AJ....115.1693C} and Faint Images of the Radio Sky at Twenty centimeters \citep[FIRST,][]{1995ApJ...450..559B} catalogs within 1 arcsec of their optical positions. We found that 36 sources (6\%) have radio counterparts (with signal-to-noise ratio $>9$) in the FIRST survey. Not surprisingly, NGC~4395 \citep[][]{2006ApJ...646L..95W} and RGG\,9 \citep[][]{2020MNRAS.495L..71Y} are among these sources but are excluded from the present sample, since these sources have already been observed in VLBI. We further made a selection using the following criteria: (1) an estimated black hole mass from literature of $< 10^6 M_\odot$, (2) a FIRST radio flux density $>2$ mJy (signal-to-noise ratio $>12$), (3) sources which are X-ray detected, providing additional support for the presence of an AGN. Finally, four sources satisfy our selection criteria, see Table \ref{tab:info} for their main properties. The IMBH candidates in our sample are RQ-AGN ($\mathcal{R} < 10$) with $\log R_{\rm X}$ in the range $-5.6$ to $-3.9$ and Eddington ratios $\log \lambda_{\rm Edd}$ in the range $-1.4$ to $-0.5$ (sub-Eddington accretion sources).

\begin{deluxetable*}{cccccccccc}
\tablecaption{IMBH candidates involved in our observational campaign. \label{tab:info}}
\tablewidth{0pt}
\tablehead{
\colhead{Alias} & \colhead{SDSS} & \colhead{$z$} & \colhead{$\log{M_\mathrm{BH}}$} & \colhead{$\log{L_\mathrm{X(2-10keV)}}$} & \colhead{$\log{L_\mathrm{H\alpha}}$} & \colhead{$\log{L_\mathrm{B}}$} & \colhead{$\log{\lambda_\mathrm{Edd}}$} & $\log{\mathcal{R}}$ & $\log{R_\mathrm{X}}$\\
\colhead{} & \colhead{} & \colhead{}& \colhead{($M_\odot$)}& \colhead{(erg\,s$^{-1}$)}  & \colhead{(erg\,s$^{-1}$)} & \colhead{(erg\,s$^{-1}$)} & & &
}
\decimalcolnumbers
\startdata
GH047            & J082443.28$+$295923.5   &$0.025$&$ 5.6^\dagger  $&$ 42.4^a, 42.5^b         $&$40.3$&$41.9$&$-0.8$&$-0.01$&$-5.6$ \\
GH106            & J110501.98$+$594103.5   &$0.033$&$ 5.5^\dagger  $&$ 42.1^a                 $&$40.5$&$42.1$&$-0.5$&$0.85$&$-4.2$ \\
GH158            & J131659.37$+$035319.9   &$0.045$&$ 5.8^\dagger  $&$ 41.7^a                 $&$40.6$&$42.2$&$-0.7$&$0.71$&$-3.9$ \\
GH163            & J132428.24$+$044629.6   &$0.021$&$ 5.7^{\dagger}, 5.8^{\star} $&$ 41.7^a $&$39.7$&$41.4$&$-1.4$&$0.80$&$-4.6$ \\
\hline
\enddata
\tablecomments{Columns give (1) identification from \citet{2007ApJ...670...92G}, (2) SDSS name, (3) redshift, (4) black hole mass
(5) X-ray luminosity, (6) H$\alpha$ line luminosity, obtained from papers where black hole masses were measured, (7) $B$-band luminosity, estimated from H$\alpha$ line luminosity \citep[see][]{2007ApJ...670...92G, 2012ApJ...755..167D, 2020ApJ...904..200Y}, and (8-10) Eddington ratio, radio loudness and radio to X-ray luminosity ratio, which are defined as $\lambda_\mathrm{Edd}{\equiv}L_\mathrm{bol}/L_\mathrm{Edd}$, $\mathcal{R}{\equiv}1.3\times10^5L_\mathrm{5GHz}/L_\mathrm{B}$ and $R_\mathrm{X}{\equiv}L_\mathrm{5GHz}/L_\mathrm{X}$, respectively, where $L_\mathrm{bol}=10L_\mathrm{B}$ and $L_\mathrm{Edd}=1.26\times10^{38}(M_\mathrm{BH}/M_\odot)$\,(erg/s) \citep[see also][]{2020ApJ...904..200Y}. Here we take 5\,GHz luminosity estimated from VLBA L-band (see Table \ref{tab:prop}).}
\tablecomments{References for black hole mass. $\dagger$: \citet{2007ApJ...670...92G}; $\star$: \citet{2012ApJ...755..167D}.}
\tablecomments{References for X-ray luminosity. $a$: \citet{2014ApJ...788L..22G}; $b$: \citet{2014AA...569A..71C}; 
}
\end{deluxetable*}

\section{Observation and data reduction}

The VLBA observations were conducted from May 31 to July 10, 2021 (UT) under the project BA146 (see Table \ref{tab:log} for more information). The observations were scheduled at L-band (the central frequency is 1.545\,GHz, hereafter we will use 1.5\,GHz, in short), with a total observation time of 12 h and a data recording rate of 2\,Gbits per second. Phase-referencing mode was used, and a nearby ($<2^\circ$) compact and strong radio source was chosen as a phase-reference calibrator for each target (see Table \ref{tab:log}). The correlated data were processed using the Astronomical Image Processing System \citep[AIPS,][]{2003ASSL..285..109G} developed by the National Radio Astronomy Observatory (NRAO) of the USA. Apriori amplitude calibration was performed using the system temperatures and antenna gain curves provided by each VLBA station. The Earth orientation parameters were obtained and calibrated using the measurements from the U.S. Naval Observatory database, and the ionospheric dispersive delays were corrected from a map of the total electron content provided by the Crustal Dynamics Data Information System (CDDIS) of NASA \footnote{\url{https://cddis.nasa.gov}}. The opacity and parallactic angles were also corrected using the auxiliary files attached to the data. The instrumental delay in the visibility phase was calibrated using a strong fringe-finder source. Finally, a global fringe fitting on the phase-reference calibrator was performed, taking the calibrator's phase model to solve miscellaneous phase delays of the target.

\begin{figure*}
\centering
\includegraphics[scale=0.6]{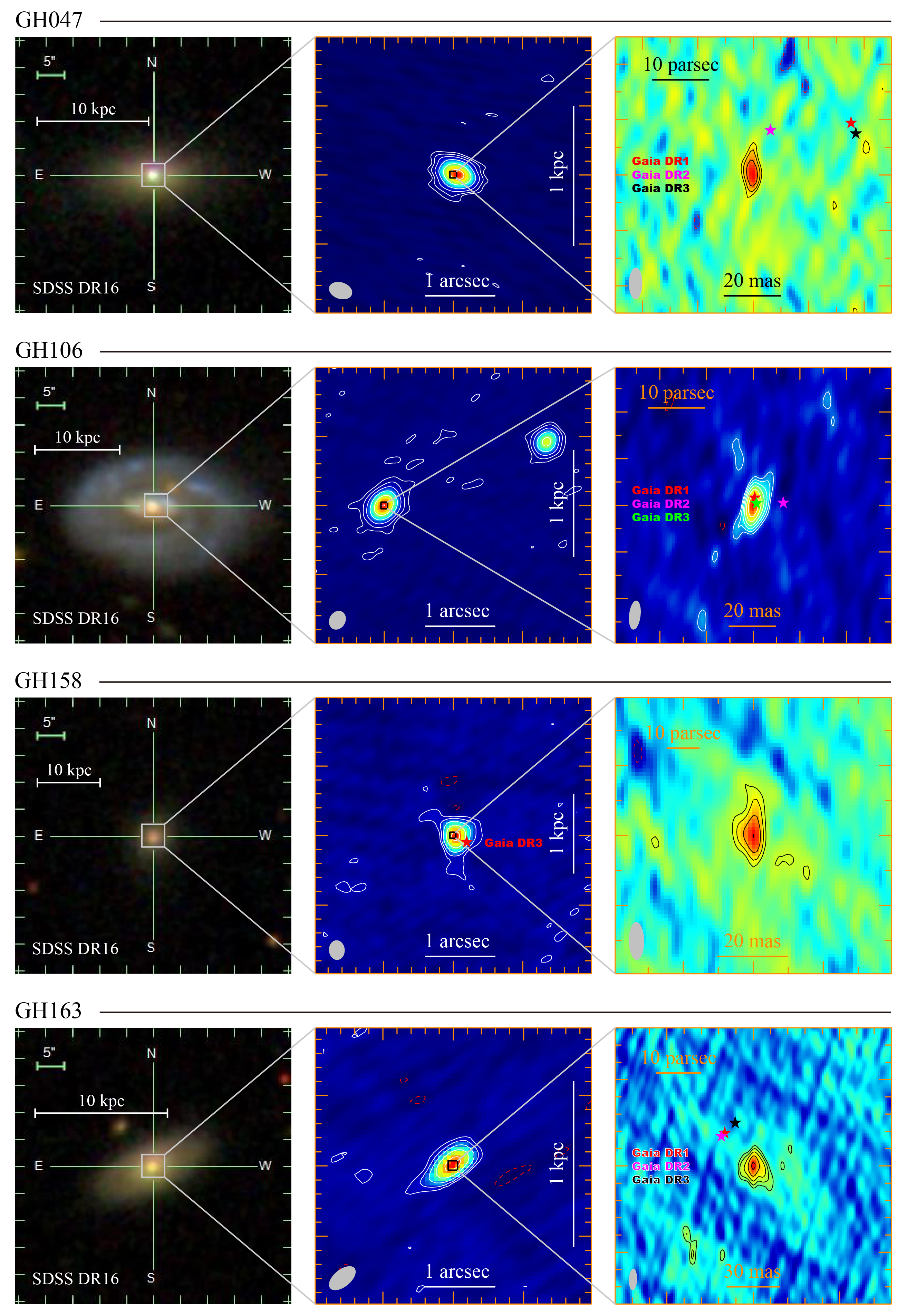}
\caption{\footnotesize \textbf{Multi-band images of the 4 IMBH candidates with core detected}. Here the image matrix is collected as one source one line, where the left column is from SDSS Data Release 16; the middle column is from VLA A-array X-band (9\,GHz) and the right column is from VLBA L-band (1.5\,GHz). Each map is centered on the VLBA peak (see Table \ref{tab:results}), except the VLA image of GH106. The black and white solid contours represent positive values and the red dashed contours represent negative values. The contours are at 3$\sigma\times(-1, 1, 2, 4, 8,...)$ for VLA images, while the contours are at 3$\sigma\times(-1, 1, 1.41, 2, 2.83,...)$ for VLBA images. Here $1\sigma$ noises are 0.016, 0.061, 0.031, 0.025\,mJy\,beam$^{-1}$ for VLBA L-band images from the top to the bottom, respectively, and they equal to the uncertainty of peak flux density (see Table \ref{tab:results}) divided by $1.8$ for VLA images. The grey ellipses in the bottom left corner of each panel represent the full width at half-maximum (FWHM) of the restoring beam (see Table \ref{tab:results}). Markers in the middle and right column are the optical coordinates obtained from \textit{Gaia} data release 1 (DR1), data release 2 (DR2) and data release 3 (DR3), where the astrometric uncertainty for \textit{Gaia} is too small to be marked. \label{fig:map}}
\end{figure*}

\begin{figure*}
\centering
\includegraphics[scale=0.5]{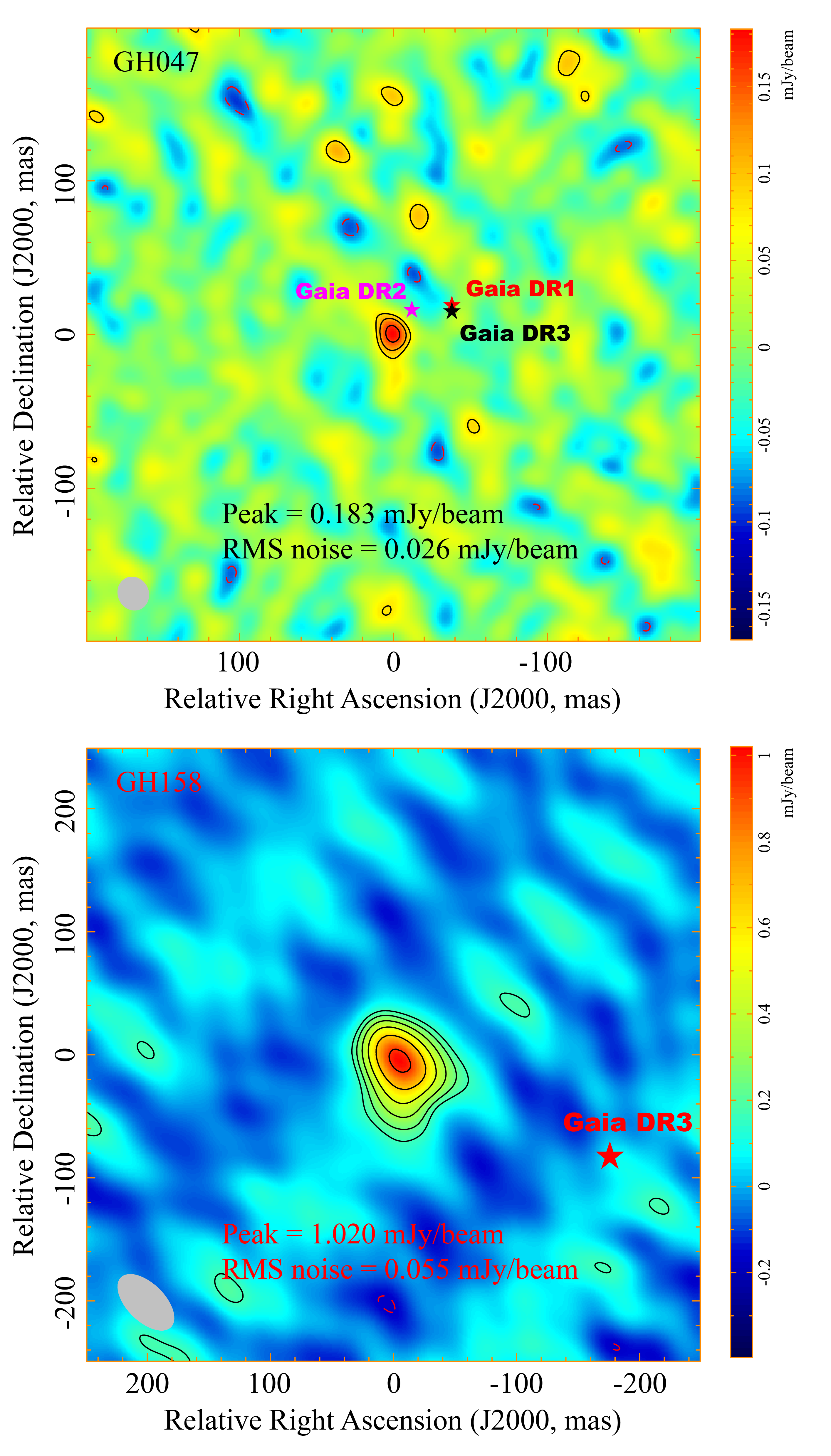}
\caption{\footnotesize \textbf{UV-tapered VLBA 1.5\,GHz images of GH047 and GH158}. The black solid contours represent positive values and the red dashed contours represent negative values. The contours are at 3$\sigma\times(-1, 1, 1.41, 2, 2.83,...)$. The peak flux densities and $1\sigma$ noises are marked in the images. The grey ellipses in the bottom left corner of each panel represent the full width at half-maximum (FWHM) of the restoring beams, they are $22.4\times20.1$ (mas) at a position angle of $22.4^\circ$ and $57.4\times30.9$ (mas) at a position angle of $45.2^\circ$ for GH047 and GH158, respectively. \label{fig:tap}}
\end{figure*}

The calibrated data of targets were exported into DIFMAP \citep{1997ASPC..125...77S} for imaging and model fitting. The final images were created using natural weights, see right panel of Figure \ref{fig:map}. Both GH047 and GH158 have the signal-to-noise ratio of $\sim8$ in the full resolution VLBA 1.5\,GHz images. We performed a uv-taper in DIFMAP to identify the detection in GH047 and GH158, see Figure \ref{fig:tap}, where GH158 shows a clear detection with a signal-to-noise ratio of $18$, while GH047 still has a signal-to-noise ratio of $\sim7$ and indicates a weak and compact radio emission.

For VLBA data, we estimate flux density uncertainties following the prescription of \citet{1999ASPC..180..301F}. The integrated flux densities $S_i$ are extracted from Gaussian model-fit in DIFMAP with the task `MODELFIT', where a standard deviation in model-fit is estimated for each component and considered as the fitting noise error. Additionally, we assign a standard 5\% error originating from amplitude calibration of VLBA (see VLBA Observational Status Summary 2021A \footnote{\url{https://science.nrao.edu/facilities/vlba/docs/manuals/oss2021A}}).

Furthermore, we obtained VLA data of our targets from the NRAO archive, which were observed earlier under the two projects AG0777 (PI: Joan Wrobel; observed in 2008), and, 12B-064 plus SD0129 \citep[see][]{2014ApJ...788L..22G} (observed from the end of 2012 to the beginning of 2013). No image has been published from any of the archival data, and hence, we manually reduced the data using the Common Astronomy Software Application \citep[CASA v5.1.1,][]{2007ASPC..376..127M} following the procedure described in \citet{2020ApJ...904..200Y}. The model-fitting results of the VLA data are listed in Table \ref{tab:results}. Here, the uncertainties on the integrated and peak flux density of the VLA data are estimated using the method described in \citet{2020ApJ...904..200Y}.

\section{Results}
In the VLBA 1.5\,GHz observations, all four sources are detected with a signal-to-noise ratio above 5$\sigma$. Figure \ref{fig:map} shows the SDSS DR16, the VLA A-array 9\,GHz, and the VLBA 1.5\,GHz images for each source (from left to right). The VLA A-array 1.4 and 9\,GHz and VLBA 1.5\,GHz observational results are listed in Table \ref{tab:results}. The VLA A-array X-band observations used a wide-band filter with a bandwidth of 2\,GHz (from 8\,GHz to 10\,GHz). The estimated in-band spectral indices between 8.5 and 9.5\,GHz are listed in column 8 of Table \ref{tab:prop}. 

In the VLA A-array X-band image of GH106, a nearby ($\sim3$\,arcsec away from the core region, R.A.=11$^h$05$^m$01$^s$.6734, Dec.=59$^\circ$41$^\prime$04$^{\prime\prime}$.430, J2000) radio source is detected in our data analysis (see Figure \ref{fig:map}), which is located in the nuclear region of the host galaxy. The component has a $8.5-9.5$\,GHz in-band spectral index of $-0.10$, implying a flat spectrum. With limited sensitivity and resolution, VLA A-array L-band observations have only marginally detected the component, the VLA A-array L/X spectral index is $-0.34$, which still implies a flat spectrum. The component is not detected in our VLBA L-band observations, and hence, it is less compact than the nucleus. Based on the flat radio spectrum and the size, this component is possibly a ($<$0.3\,arcsec or $<$200\,parsec in size) HII region \citep[see e.g.][]{1997ApJ...488..621U, 1997ApJS..109..417L}.

At the VLBA mas-scales, all sources have compact emissions. The concentration indices (ratio of the peak $S_p$ to integrated $S_i$ flux densities) in Table \ref{tab:prop} are indicative of GH047 and GH106 being relatively more compact than GH158 and GH163 at VLBA scales. As the VLA L-band captures emission from a larger area in comparison to the VLBA L-band observations, we can estimate the fraction of extended emission for the AGN (from milli-arcsec to arcsec scales, see column 9 of Table \ref{tab:prop}). A major fraction ($\gtrsim50\%$) of the emission is from extended scales in all AGN indicating that the VLBA observations tend to capture only the most compact emission. Interestingly, GH047 tends to have a larger fraction of arcsec-scale radio emission than the three other sources. We estimate the brightness temperature using 
\citep[e.g.][]{2005ApJ...621..123U} 
\begin{equation}\label{eq:bt}
T_\mathrm{B}=1.8\times10^9(1+z)\frac{S_i}{\nu^2\phi^2}~\mathrm{(K)},
\end{equation}
where $S_i$ (mJy) are the integrated flux densities of each Gaussian component with a full width at half maximum $\phi$ (mas). These parameters were estimated by fitting two-dimensional Gaussian models to the {\it UV} data. $\nu$ is the observing frequency in GHz, and $z$ is the redshift. The measured flux densities $S_i$, FWHM of beam $\phi$, and estimated brightness temperatures for the sources (at each observation frequency $\nu$) are presented in Table \ref{tab:results}. We note that the measured component sizes are upper limits, and therefore, the radio brightness temperatures estimated here should be considered as lower limits.

\section{Discussion}

\subsection{The fundamental plane of black hole activity}
\label{fprelation}

An empirical correlation among black hole mass, radio, and X-ray luminosity spanning XRBs (hosting stellar-mass black holes) and AGN (hosting SMBHs) is termed the fundamental plane relation of black hole activity \citep{2003MNRAS.345.1057M}. This relation may be affected by radio emission produced in a jet/relativistic outflow, X-ray emission produced in a disk-corona system, and if both the radio and X-ray power are related to the black hole mass and accretion rate \citep[see, e.g.][]{2017SSRv..207....5R}. Therefore, the fundamental plane relation may be applicable to any accretion-powered system during a low/hard state \citep{2006A&A...456..439K}, or in an intermediate state that can involve the production of episodic/intermittent radio ejecta \citep{2004MNRAS.355.1105F}.

\begin{figure*}
\centering \includegraphics[scale=0.7]{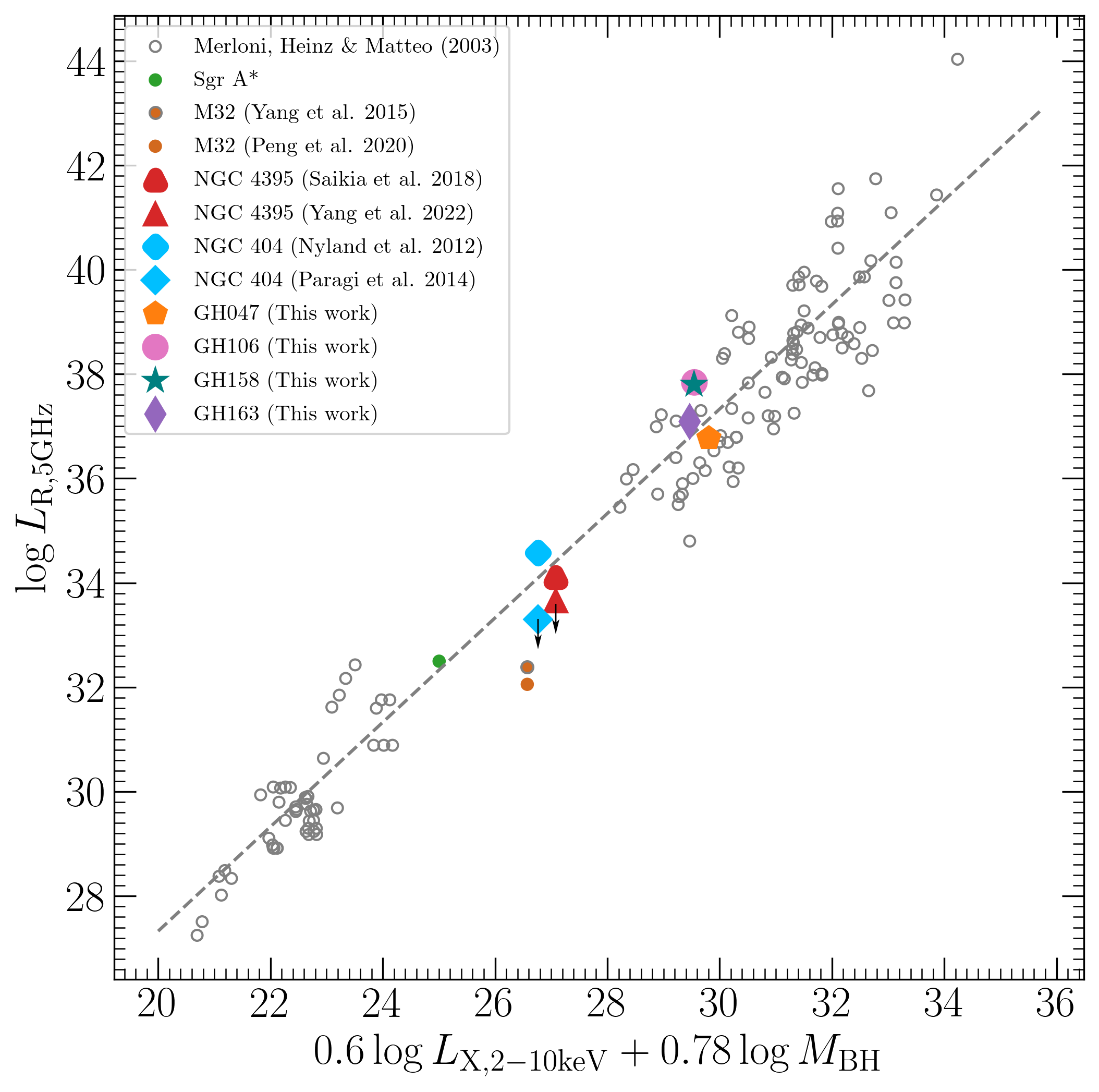}
\caption{The fundamental plane of black hole activity based on \citet{2003MNRAS.345.1057M}. The references in the legend show where the radio luminosity was taken. The black open squares and data for Sgr A$^\ast$ are from \citet{2003MNRAS.345.1057M}. Note that the sub-parsec-scale radio luminosity (obtained from EVN observations) for NGC~4395 (Yang et al. In preparation) and NGC~404 \citep[][]{2014ApJ...791....2P} are only upper limits. \label{fig:fmp}}
\end{figure*}

The validity of the fundamental plane relation for the current sample of potential IMBH hosting AGNs can be tested with the available X-ray and radio luminosities, and the measured black hole masses (see Table \ref{tab:info}). The scatter in the relation can be contributed by the non-nuclear and extended radio emission \citep[e.g.][]{2018MNRAS.477.2119S}. For the IMBH candidates in this work, the VLA-based radio flux densities at arcsec resolutions probe the kpc-scales, which can include contributions from the host galaxy. The radio emission for the AGN components may thus be overestimated if including the extended emission. Employing only the radio luminosity estimated from VLBA L-band flux densities, and the previously inferred X-ray luminosities and black hole masses (see Table \ref{tab:info}), the IMBH candidate sample is found to closely follow the fundamental plane relation (see Figure \ref{fig:fmp}). VLBA observations of a sample of RQ-AGNs indicate that for sources with $\lambda_{\rm Edd} \lesssim 0.3$, the radio emission originates from a compact region, potentially the size of the accretion disk \citep{2022ApJ...936...73A}; such a scenario may be operational in the current sample. For scale-independent physics, a similarity with XRB spectral state transitions places these IMBH sources in a high-soft state, characterized by a dominant accretion-based emission. The four sources in the sample have Eddington ratios of $0.04 - 0.32$ (see Table \ref{tab:info}) indicating a prominence of accretion-powered activity and the relative suppression of a radio-emitting jet \cite[e.g.][]{2003MNRAS.344...60G}. The transition from a low-hard to a high-soft state involves the ejection of material (at relativistic velocities) that is likely to be responsible for the production of radio emissions. The pc-scale outflow/jet may possibly be a signature of the above process and can be similar to the discrete ejection event observed in the IMBH candidate ESO 243-49 HLX-1 \citep{2012Sci...337..554W}.

We include three additional low-luminosity AGNs, two of which are potential IMBH hosts. The availability of radio flux densities (and corresponding radio luminosity) from their pc-scale nuclear region, X-ray luminosities, and black hole mass estimates can be used to test their position on the fundamental plane relation, in an intermediate region (between the stellar mass and supermassive scales) that is less populated. The sources include NGC 4395, NGC 404, and M32. For NGC 4395, the Eddington ratio is $1.2 \times 10^{-3}$ \citep{2006ApJ...646L..95W}, and the black hole mass is $\approx 10^4~M_\odot$ \citep{2019NatAs...3..755W}. The 5 GHz radio luminosity of $1.3 \times 10^{34}$ erg s$^{-1}$ is estimated from the VLA A array 15 GHz observations \citep{2018A&A...616A.152S} which probes a region size of 4 pc \citep[taking the in-band 15\,GHz spectral index of $-0.07$, see also][]{2022MNRAS.514.6215Y}, and the X-ray luminosity is $10^{39.9}$ erg s$^{-1}$ \citep{2005AJ....129.2108M}. For NGC 404, the Eddington ratio is $1.5\times 10^{-6}$ \citep{2014ApJ...791....2P}, and the black hole mass $\approx 5 \times 10^5~M_\odot$ \citep{2020MNRAS.496.4061D}. The radio luminosity is $3.2 \times 10^{34}$ erg s$^{-1}$ based on the VLA A array 5 GHz observations \citep{2012ApJ...753..103N} and $<2 \times 10^{33}$ erg s$^{-1}$ based on EVN 5 GHz observations \citep{2014ApJ...791....2P} that probe the sub-pc  -- pc scale region, and the X-ray luminosity is $1.4 \times 10^{37}$ erg s$^{-1}$ based on {\it Chandra} observations \citep{2014ApJ...791....2P}. The source M32 is included here despite having a black hole mass of $\approx 3 \times 10^6~M_\odot$ mainly owing to a low Eddington ratio of $\approx 3.2 \times 10^{-9}$ \citep{2003ApJ...589..783H}. The radio luminosity is $2.5 \times 10^{32}$ erg s$^{-1}$ based on VLA B array 6.6 GHz observations \citep{2015ApJ...807L..19Y} and $10^{32}$ erg s$^{-1}$ based on VLA A array 6 GHz observations \citep{2020ApJ...894...61P} that probe regions of size $\approx$ 4 pc and $\approx$ 1.5 pc respectively. The X-ray luminosity is $7.9 \times 10^{35}$ erg s$^{-1}$ based on {\it Chandra} observations \citep{2003ApJ...589..783H}.

These sources are also found to follow the fundamental plane relation (see Figure \ref{fig:fmp}). The low Eddington ratios ($\leqslant 10^{-3}$) for these sources are indicative of a radiatively inefficient mode of accretion onto the central black hole \cite[e.g.][]{2008ARA&A..46..475H,2009ApJ...699..626H,2014ARA&A..52..529Y}. This accretion mode is found to generally characterize a large sample of RQ-AGNs that follow the fundamental plane relation \cite[e.g.][]{2022MNRAS.513.4673B}. A major portion of the gravitational binding energy and angular momentum accrued by the accreting radiatively inefficient gas in the vicinity of the black hole can be transported by global winds or outflows \cite[e.g.][]{1999MNRAS.303L...1B,2000MNRAS.311..507D,2008ARA&A..46..475H}. The fundamental plane relation thus captures this activity regime and is indicative of a low-hard spectral state if similar to the XRBs \cite[e.g.][]{2004A&A...414..895F}. A moderate resolution (corresponding to physical scales of a few pc) and high sensitivity VLA (for nearby sources, for example, NGC~4395, NGC\,404, and M32) or VLBI observation (for sources in this work) can thus help capture the outflow/ejection activity of the central engine in the hard to the intermediate spectral regime.

\subsection{The nature of radio emission}

All four sources have steep radio spectra between VLA A-array L and X-band (between $-0.63$ to $-1.05$), and span angular scales between $\approx 0.2$ to $2$\,arcsec (physical size of tens of pc to kpc, see Tables \ref{tab:results} and \ref{tab:prop}). Comparing the spectral indices as inferred from the VLA L/X band and only from the X-band (in-band), the spectra of GH047 and GH158 tend to become steeper while that of GH106 and GH163 tend to become flattered. This implies the prevalence of older/relic emissions in GH047 and GH158 compared to GH106 and GH163, where the emission is more recent. However, the concentration index of the VLA X-band emission exceeds or is comparable to that in the VLA L-band emission in all sources (see Table \ref{tab:prop}), suggesting the emergence of new emission from a flat spectrum compact component at higher frequencies. The in-band indices in the X-band (8.5 to 9.5 GHz) probe regions of size $\approx$ 100 pc, indicate that the recent activity in GH106 and GH163 is from the compact scales. The above picture is consistent with the coincidence of the radio emission peak positions in their VLA 9\,GHz and VLBA 1.5\,GHz images, and indicating an active pc-scale region (see Figure \ref{fig:map}). Though with low signal-to-noise ratio ($\sim4-6$), there are possibly emergent outflow structures along the NW-SE direction in GH106 and E-W direction in GH163 (see Figure \ref{fig:map}) that may allude to the above picture. Again, the flat spectrum (in-band index of $-0.26 \pm 0.17$ at 9 GHz) in GH106 indeed indicates the prevalence of optically thick emission in the nuclear region.

The {\it Gaia} mission is primarily aimed at measuring the spatial position (astrometry) and velocity information from Galactic stars through photometric and spectroscopic surveys \citep{2016A&A...595A...1G}. The instrument is suitable for VLBI-like measurements of the astrometric information for background non-stellar sources including AGNs that may be present in the field of view \cite[e.g.][]{2019MNRAS.482.1701Y}, with sub-mas astrometric uncertainties for bright sources \citep{2016A&A...595A...2G, 2018A&A...616A...1G, 2021A&A...649A...1G}. The {\it Gaia} optical and VLBA L-band radio positions for GH047 and GH163 are offset by $\approx$ 20 - 30 mas, that of GH106 are coincident and that of GH158 is offset by $\sim$ 0.15 arcsec (see the marker in Figure \ref{fig:tap} and VLA X-band image of Figure \ref{fig:map}). An offset itself may be an indicator that the {\it Gaia} observations track the accretion activity while the VLBI observations track the outflow activity. The relatively smaller offset ($\leqslant$ 20 mas) or near coincidence for GH106 and GH163 are in agreement with the accretion-outflow activity being more recent in these sources.

The radio emission in these sources at the compact sub-pc to pc scales may originate from corona mass ejection or winds from the accretion disk \cite[e.g.][]{2019NatAs...3..387P, 2020ApJ...904..200Y}. This is indicated by the radio VLBI to X-ray luminosity ratio $(L_{\rm 5~GHz}/L_{\rm 2-10~keV})$ for the four sources that span between $2.5 \times 10^{-6}$ to $1.3 \times 10^{-4}$. This spans a regime similar to that in coronal active stars \citep{1993ApJ...405L..63G} effected in the presence of strong magnetic fields (e.g. re-connection events) that can produce coronal mass ejection \cite[e.g.][]{2008MNRAS.390..847L,2019NatAs...3..387P}.

The non-thermal nature of the large-scale radio emission, steep spectral indices (between the VLA L and X band observations), and morphology suggest an un-beamed extended emission region. Further, the emission is attributable to the extended structure dominating over that from the compact VLBI scales as inferred from the measured flux densities. The growth of the pc-scale outflow to the large-scale structure may proceed through intermittent/episodic activity \cite[e.g.][]{2020ApJ...905...74N} for a scenario similar to the XRBs \cite[in the low-hard state or a transition from low-hard to high-soft X-ray spectral shape, e.g.][]{2009MNRAS.396.1370F} as has been inferred from Section \ref{fprelation}. A coupling of the outflow with the accretion activity entails an intermittent/episodic nature \cite[e.g.][]{2009ApJ...698..840C}. The kpc-scale morphology can be structured by emission from a past jet ejection \cite[e.g.][]{2009ApJ...698..840C,2020ApJ...905...74N}, and follows a growth trajectory similar to young AGN or peaked spectrum sources \cite[e.g.][]{2012ApJ...760...77A,2016MNRAS.459.2455C,2021A&ARv..29....3O}. In this scenario, the radio power can increase (scaling with time as $\propto t^{2/5}$) governed only by adiabatic losses; following this stage, synchrotron losses begin to dominate as the source grows to the kpc-scale, with the power flattening, both stages accompanied by steep (index of $\approx - 1.0$) spectra \cite[e.g.][]{2007MNRAS.381.1548K,2012ApJ...760...77A}. This may explain the relative dominance of the extended-scale emission over that from the pc-scale outflow.

Physical properties of the emitting region can then be estimated by modeling the synchrotron emission that ensues from electrons accelerated by the expanding shock. Assuming a power-law electron number density distribution $N(\gamma) = K \gamma^{-p}$ where $p$ is the index related to the optically thin spectral index $\alpha$ as $p = 1-2 \alpha$, and $K$ is the normalization, the optically thin flux density (integrated over the pitch angle) can be expressed in terms of the synchrotron emissivity $j_\nu$ as \citep{1998ApJ...499..810C,2013LNP...873.....G}
\begin{equation}
F_\nu = j_\nu \frac{4 \pi R^3 f_V}{3 D^2_L},\label{Fnu}
\end{equation}
In the above expression, $f_V \leqslant 1$ is the volume filling factor in a spherical region of size $R$, and $D_L (z)$ is the redshift dependent distance to the source \cite[e.g.][]{1999astro.ph..5116H} and, the emissivity $j_\nu$ is expressed in terms of the particle normalization $K$, the magnetic field energy density in the region $U_B = B^2/(8 \pi)$ (where $B$ is the magnetic field strength), and the energy index $p$ as \cite[eqn. 4.45,][]{2013LNP...873.....G}
\begin{equation}
j_\nu = \frac{3 \sigma_T c K U_B}{16 \pi^{3/2} \nu_L} \left(\frac{\nu}{\nu_L}\right)^{-(p-1)/2} f_j (p),\label{jnu}
\end{equation}
where $\sigma_T = 6.65 \times 10^{-25}$ cm$^2$ is the Thomson cross section for electrons, $c = 3.0 \times 10^{10}$ cm s$^{-1}$ is the speed of light, $\nu_L = e B/(2 \pi m_e c)$  is the Larmor frequency (where $e = 4.80 \times 10^{-10}$ esu is the unit electric charge, and $m_e = 9.11 \times 10^{-28}$ g is the electron mass), and $f_j (p)$ in eqn. 4.46 of \citet{2013LNP...873.....G}.
The emission is from a region of size $R$ and a volume filling factor $f_V \leqslant 1$. The distance to the source $D_L (z)$ is given by using $z$ from Table \ref{tab:info} and assuming a standard $\Lambda$CDM with parameters given in Section \ref{sec:intro}. Assuming an equipartition of the total energy density between the particle kinetic energy and magnetic fields, we obtain the following relation between the normalization $K$ and the magnetic field strength $B$ \citep{1998ApJ...499..810C,2020ApJ...903..116A}
\begin{equation}
K = \frac{\epsilon_e}{\epsilon_B} \frac{B^2}{8 \pi} (p-2) \frac{\gamma^{p-2}_{\rm min}}{m_e c^2},\label{Kexp}
\end{equation}
where $\epsilon_e$ and $\epsilon_B$ are microphysical parameters (in addition to $p$) and represent the fractions of the total energy density (of the shocked material) in the particle kinetic energy density and in the magnetic field respectively, and ${\displaystyle \gamma_{\rm min} \approx \left(\frac{p-2}{p-1}\right) \epsilon_e \left(\frac{m_p}{m_e}\right)}$ \cite[e.g.][]{2013NewAR..57..141G} is the minimum Lorentz factor of injected electrons (where $m_p = 1.67 \times 10^{-24}$ g is the proton mass). The eqns. (\ref{jnu}) -- (\ref{Kexp}) are used in the expression for the flux density in eqn. \ref{Fnu}. The resulting equation can be inverted to express the magnetic field strength in a parametric form 
\begin{align}
B &= \left(\frac{128 \pi^{3/2} e}{\sigma_T} \left(\frac{e}{2 \pi m_e c}\right)^{-(p-1)/2}\right)^{2/(p+5)} \\ \nonumber  & \left(\frac{F_\nu~D^2_L~\nu^{(p-1)/2}}{(\epsilon_e/\epsilon_B)~f_V~f_j(p)~(p-2)~\gamma^{(p-2)}_{\rm min} R^3}\right)^{2/(p+5)}. 
\end{align}
The total energy in the region can be evaluated as
\begin{equation}
E = \frac{4 \pi}{3} R^3 f_V \frac{U_B}{\epsilon_B} = \frac{f_V}{6 \epsilon_B} B^2 R^3.
\end{equation}
The above basic framework is independent of IMBH mass, making it applicable to a range of compact object systems hosting intermittent/episodic or sustained outflow activity. 

The magnetic field strength $B$ and minimum total energy $E$ in the emitting region (resolved, extended) are evaluated for the potential IMBH hosts. The flux density $F_\nu = S_{i} - S_{p}$ with the VLA L-band values for $S_{i}$ and $S_{p}$ taken from Table \ref{tab:results}. The region size $R$ is evaluated based on a weighting involving the concentration index (see Table \ref{tab:prop}), with $R = R_{min,{\rm VLA}} (1-S_p/S_i)-R_{min,{\rm VLBA}}$. The evaluated $R$ range is between 33.8 pc (GH047) -- 350.2 pc (GH106); this region is likely to span between the VLBA pc-scale and the VLA sub-kpc to kpc scales. We use an observational frequency $\nu = 5.0$ GHz, a power law index $p = 1- 2 \alpha$ (using the L/X band spectral indices reported in Table \ref{tab:results}) and the assumptions $f_V = 0.5$, and $(\epsilon_e, \epsilon_B)$ of $(1/3,1/3)$. The magnetic field strength $B$ ranges between 0.06 -- 0.13 mG, and the total energy ranges between $0.05 - 12.14 \times 10^{53}$ erg, both values range subject to the free parameters in the model. The input parameters and the above estimates are tabulated in Table \ref{tab:largescaleemission}. The above $B$ values are similar to the mG estimates for the kpc-scale radio cores of radio-intermediate AGN \cite[e.g.][]{2021MNRAS.507..991S}. Imaging the sources at lower frequencies and in polarization may potentially unravel faint emission structures including kpc-scale lobes where $\mu$G magnetic field strengths are expected \cite[e.g.][]{2022MNRAS.513.4208S}.

\section{Summary and conclusions}
The presence of AGN in dwarf galaxies \cite[relatively isolated, mostly secular evolution,][]{2017IJMPD..2630021M,2020ARAA..58..257G} provides an opportunity to study the accretion - jet/outflow activity powered by a potential IMBH central engine. Synchrotron emission can be produced by physical processes involving an outflow; the evolutionary phase in these AGN systems may be similar to that in XRBs (spectral state transitions) owing to a possible scale invariance of the underlying physical processes.  

The above mechanisms of emission and evolution from compact to large scale (pc - kpc) structure are investigated through a set of radio observations spanning multiple resolutions (and physical scales) involving four potential IMBH hosting AGNs. This includes our VLBA L-band (1.4\,GHz) high-resolution observations (milli-arcsecond scale; probing the few to tens of pc) conducted in 2021, and archival VLA A-array L and X-band (1.5 GHz and 9 GHz respectively) intermediate to larger resolution observations (sub-arcsec to arcsec scales; probing $\approx$ hundred pc to kpc). At the VLA scales, the source is compact and unresolved, though with a dominant emission. At the VLBA scales, the sources become marginally resolved to allow the discerning of extensions from the compact emission regions.

The validity of the empirical fundamental plane of black hole activity \citep{2003MNRAS.345.1057M} is tested for a sample of sources including the four putative IMBH candidates and three additional low luminosity (and/or intermediate-mass) AGNs. As these sources span a wide range of observables (radio and X-ray luminosities, and BH mass), they can help discern the role of accretion - jet activity in affecting the fundamental plane relation. Including only the pc-scale region contribution to the radio emission (probing an active or recent outflow), we find that all sources are accommodated on the relationships within the general scatter and populate a region intermediate to that occupied by the XRBs and AGNs (see Figure \ref{fig:fmp}). 

In comparison to the XRB spectral states, the IMBH candidates are likely in the high-soft state (accretion-dominated emission) indicating that the outflow may have been recently ejected (from a low-hard outflow-dominated state or an intermediate state). The low luminosity AGNs are likely in the low-hard state involving ongoing outflow activity. The fundamental plane relation may thus be capturing the outflow and ejecta activity spanning the low-hard and a recent intermediate spectral state.

The radio emission from the compact pc-scales may be sourced as corona mass ejection or disk winds \cite[e.g.][]{2008MNRAS.390..847L,2019NatAs...3..387P}, as indicated by the relative strength of the radio luminosity in comparison with the X-ray luminosity (that spans the range of $\approx 10^{-6} - 10^{-4}$), a regime similar to that in coronal active stars \citep{1993ApJ...405L..63G}, and at high accretion rates \cite[e.g.][]{2020ApJ...904..200Y}. An offset in the emission center between the radio VLBI and optical {\it Gaia} is indicative of the former tracking the outflow, and the latter tracking the accretion. A relatively smaller offset ($\leqslant$ 20 mas) or coincidence of the positions is indicative of the outflow being recently ejected.

Emission from the kpc-scales is found to dominate over that from the pc-scale, possibly from past episodic/intermittent ejections that follow a trajectory of growth similar to that in young AGN and peak spectrum sources \cite[e.g.][]{2012ApJ...760...77A,2021A&ARv..29....3O}. In this scenario, the luminosity tends to increase with time and then reaches a plateau phase when the growing structure reaches the kpc-scale \cite[e.g.][]{2007MNRAS.381.1548K,2012ApJ...760...77A}. The resultant synchrotron emission from the intermediate scales of a few ten pc to kpc is modeled to estimate a magnetic field strength of $0.06 - 0.13$ mG and total energy of $0.05 - 12.14 \times 10^{53}$ erg. These point to a relatively less powerful central engine in the dwarf galaxies (radio emission in comparison with the optical and X-ray emission) when compared to AGNs powered by SMBHs, with potentially differing growth mechanisms of the large-scale structure. 

High to moderate-resolution VLBI observations can probe the sub-pc to the pc-scale region in nearby dwarf galaxies hosting AGN. As these putatively harbor IMBHs powering the central engine, their VLBI monitoring can help in the understanding of the onset and evolution of accretion-jet activity, bridging the divide between the XRBs hosting stellar-mass BHs and AGN hosting SMBHs, and can play a complementary role in a multi-wavelength perspective.

\begin{acknowledgments}
This work is supported by the Shanghai Sailing Program (21YF1455300) and China Postdoctoral Science Foundation (2021M693267). XLY is thankful for the support from the National Science Foundation of China (12103076). LCH is supported by the National Science Foundation of China (11721303, 11991052, 12011540375), China Manned Space Project (CMS-CSST-2021-A04), and the National Key R\&D Program of China (2016YFA0400702).
XFY is supported by the CAS Pioneer Hundred Talents Program.
Scientific results from data presented in this publication are derived from the VLBA project BA146.
The National Radio Astronomy Observatory is a facility of the National Science Foundation operated under cooperative agreement by Associated Universities, Inc. 
This work has made use of data from the Pan-STARRS1 Surveys (PS1) and the PS1 public science archive. 
This work has made use of data from the European Space Agency (ESA) mission {\it Gaia} (\url{https://www.cosmos.esa.int/gaia}) processed by the {\it Gaia} Data Processing and Analysis Consortium (DPAC, \url{https://www.cosmos.esa.int/web/gaia/dpac/consortium}). 
\end{acknowledgments}

\bibliography{imbh}{}
\bibliographystyle{aasjournal}




\begin{deluxetable*}{lccccl}
\tablecaption{VLBA L-band observational logs. \label{tab:log}}
\tablewidth{0pt}
\tablehead{
\colhead{Project ID} & \colhead{Target} & \colhead{Calibrator} & \colhead{Distance} & \colhead{Date} & \colhead{Antennas}
}
\startdata
BA146A$^a$                &GH047       &J082341.1$+$292828& $0.56$ &2021-05-31  & SC-HN-NL-FD-LA-PT-KP-OV-BR-MK  \\
BA146A$^b$                &GH106       &J111013.0$+$602842& $1.02$ &2021-06-01  & SC-HN-NL-FD-LA-PT-KP-OV-BR-MK \\
BA146F$^b$                &GH158       &J131829.6$+$043010& $0.72$ &2021-07-10  & SC-HN-FD-LA-PT-OV-BR-MK \\
BA146C$^c$                &GH163       &J132626.6$+$032627& $1.42$ &2021-07-02  & SC-HN-FD-LA-PT-OV-BR-MK \\
\hline
\enddata
\tablecomments{Columns give (1) project id, (2) target alias, (3) International Celestial Reference Frame (ICRF) name of the calibrator, (4) calibrator's angular distance to the target (in degree), (5) date of the observation, (6) participating antennas.}
\tablecomments{Operators. $a$: Jessica King, $b$: Alan Kerr, $c$: Betty Ragan.}
\tablecomments{Full name of antennas. SC: St. Croix, HN: Hancock, NL: North Liberty, FD: Fort Davis, LA: Los Alamos, PT: Pie Town, KP: Kitt Peak, OV: Owens Valley, BR: Brewster, MK: Mauna Kea.}
\end{deluxetable*}

\begin{longrotatetable}
\begin{deluxetable*}{cccccccccccc}
\tabletypesize{\scriptsize}
\tablecaption{Observational results. \label{tab:results}}
\tablewidth{0pt}
\tablehead{
\colhead{Name}   &
\colhead{Date}  &
\colhead{R.A.}  &
\colhead{Dec.}  &
\colhead{$S_i$}  &
\colhead{$S_p$}  &
\colhead{$\phi$} &
\colhead{$\theta_{b,maj}$} &
\colhead{$\theta_{b,min}$}  &
\colhead{PA} &
\colhead{$\log{T_\mathrm{B}}$} &
\colhead{$\log{L_\mathrm{R}}$} \\
\colhead{} &
\colhead{}  &
\colhead{(J2000)}  &
\colhead{(J2000)}  &
\colhead{(mJy)}  &
\colhead{(mJy\,beam$^{-1}$)}&
\colhead{(mas)}&
\colhead{(mas)}&
\colhead{(mas)}&
\colhead{($\circ$)}&
\colhead{(K)}&
\colhead{(erg\,s$^{-1}$)}
}
\decimalcolnumbers
\startdata
\multicolumn{12}{c}{VLBA L-band (1.5\,GHz)} \\
\hline
GH047&2021-05-31 &08:24:43.28793 &+29:59:23.4978            &$0.185\pm0.036$&$0.135\pm0.017$&$ 4.23$&$ 11.4 $&$ 4.8 $&$ -0.45 $&$  6.9 $&$ 36.58\pm0.08^\star $\\
GH106&2021-06-01 &11:05:01.98412 &+59:41:03.5096            &$2.036\pm0.131$&$1.460\pm0.095$&$ 4.20$&$ 12.5 $&$ 4.8 $&$ -6.59 $&$  7.9 $&$ 38.39\pm0.02^\star $\\
GH158&2021-07-10 &13:16:59.38357 &+03:53:20.0327            &$0.750\pm0.080$&$0.267\pm0.033$&$ 8.62$&$ 10.8 $&$ 4.3 $&$  0.61 $&$  6.9 $&$ 38.23\pm0.04^\star $\\
GH163&2021-07-02 &13:24:28.23767 &+04:46:29.5448            &$0.836\pm0.111$&$0.325\pm0.029$&$ 9.35$&$ 12.4 $&$ 4.6 $&$ -2.23 $&$  6.8 $&$ 37.60\pm0.05^\star $\\
\hline
\multicolumn{12}{c}{VLA A-array L-band (1.4\,GHz)} \\
\hline
GH047&2008-12-01 &08:24:43.278 &+29:59:23.51 &$2.124\pm0.127$&$2.020\pm0.143$&$ 402$&$ 1850 $&$ 1440 $&$ -84.6 $&$  4.0 $&$ 38.16\pm0.02^\diamond $\\
GH106&2008-12-05 &11:05:01.998 &+59:41:03.50 &$6.525\pm0.492$&$4.100\pm0.555$&$ 1133$&$ 1730 $&$ 1450 $&$ -39.8 $&$  3.6 $&$ 38.90\pm0.03^\diamond $\\
GH158&2008-12-17 &13:16:59.381 &+03:53:20.04 &$1.855\pm0.136$&$1.790\pm0.154$&$ 56$&$ 1720 $&$ 1530 $&$ 11.2 $&$  5.7 $&$ 38.62\pm0.03^\diamond $\\
GH163&2008-12-17 &13:24:28.241 &+04:46:29.56 &$2.785\pm0.211$&$2.570\pm0.238$&$ 270$&$ 1740 $&$ 1460 $&$ 21.5 $&$  4.5 $&$ 38.13\pm0.03^\diamond $\\
\hline
\multicolumn{12}{c}{VLA A-array X-band (9\,GHz)} \\
\hline
GH047&2013-01-06 &08:24:43.2837 &+29:59:23.505 &$0.648\pm0.010$&$0.664\pm0.010$&$ 11.7$&$ 345 $&$ 239 $&$ 71.9 $&$  5.0 $&$ 37.656\pm0.006^\bullet $\\
GH106&2012-11-15 &11:05:01.9849 &+59:41:03.507 &$0.918\pm0.015$&$0.773\pm0.014$&$ 102.5$&$ 282 $&$ 231 $&$ -29.2 $&$  3.3 $&$ 38.039\pm0.007^\bullet $\\
GH158&2012-12-15 &13:16:59.3818 &+03:53:20.025 &$0.398\pm0.011$&$0.376\pm0.012$&$ 58.9$&$ 282 $&$ 225 $&$ 4.02 $&$  3.4 $&$ 37.967\pm0.013^\bullet $\\
GH163&2012-12-16 &13:24:28.2385 &+04:46:29.575 &$0.439\pm0.011$&$0.401\pm0.012$&$ 85.5$&$ 449 $&$ 242 $&$ -50.6 $&$  3.1 $&$ 37.325\pm0.011^\bullet $\\
\hline
\enddata
\tablecomments{Columns give (1) source name, (2) date of the observation, (3-4) J2000 Right Ascension and Declination coordinates, (5) integrated flux density, (6) peak flux density, (7) FWHM size of Gaussian models, (8-10) FWHM size of beam major and minor axis, and position angle, (11) brightness temperature, and (12) monochromatic luminosity.}
\tablecomments{Monochromatic luminosity. $\star$: estimated by taking the VLBA L-band radio flux density and the VLA A-array L/X spectral index; $\diamond$: estimated by taking the VLA A-array L-band radio flux density and the VLA A-array L/X spectral index; $\bullet$: estimated by taking the VLA A-array X-band radio flux density and the in-band VLA A-array wide X-band spectral index.}
\end{deluxetable*}
\end{longrotatetable}

\begin{deluxetable*}{ccccccccc}
\tablecaption{Radio observational properties of the IMBH candidates.\label{tab:prop}}
\tablewidth{0pt}
\tablehead{
\colhead{Name} & \colhead{Telescope} & \colhead{$\nu$} & \colhead{$R_{maj}$} & \colhead{$R_{min}$} & \colhead{$S_p/S_i$} & \colhead{$\log{L_\mathrm{5GHz}}$} & \colhead{$\alpha$} & \colhead{fraction} \\
\colhead{} & \colhead{} & \colhead{(GHz)}& \colhead{(pc)} & \colhead{(pc)} & \colhead{}  & \colhead{(erg\,s$^{-1}$)} & \colhead{} & \colhead{}
}
\decimalcolnumbers
\startdata
GH047            & VLBA    &$1.5$&$ 5.74  $&$ 2.41  $& $0.73\pm0.19$&$ 36.77\pm0.08^\star $&&\\
GH047            & VLA-A   &$1.4$&$ 932   $&$ 725$& 0.95$\pm$0.09&$ 37.81\pm0.03^\diamond $&$-0.63\pm0.03^{\ddagger}$&$0.90\pm0.09$\\
GH047            & VLA-A   &$9$&$   173   $&$ 120   $& 1.02$\pm$0.02&$ 38.00\pm0.03^\bullet $&$-1.38\pm0.13^{\dagger}$&\\
\hline
GH106            & VLBA    &$1.5$&$ 8.23  $&$ 3.16  $&0.72$\pm$0.06&$ 37.84\pm0.03^\star $&&\\
GH106            & VLA-A   &$1.4$&$ 1140  $&$ 955   $&0.63$\pm$0.10&$ 38.32\pm0.03^\diamond $&$-1.05\pm0.04^{\ddagger}$&$0.50\pm0.15$\\
GH106            & VLA-A   &$9$&$   185   $&$ 152   $&0.84$\pm$0.02&$ 38.10\pm0.04^\bullet $&$-0.26\pm0.17^{\dagger}$&\\
\hline
GH158            & VLBA    &$1.5$&$ 9.56  $&$ 3.80  $&0.36$\pm$0.08&$ 37.80\pm0.05^\star $&&\\
GH158            & VLA-A   &$1.4$&$ 1523  $&$ 1355  $&0.96$\pm$0.11&$ 38.17\pm0.03^\diamond $&$-0.82\pm0.04^{\ddagger}$&$0.58\pm0.10$\\
GH158            & VLA-A   &$9$&$   249  $&$ 199    $&0.94$\pm$0.04&$ 38.26\pm0.08^\bullet $&$-1.18\pm0.31^{\dagger}$&\\
\hline
GH163            & VLBA    &$1.5$&$ 5.27  $&$ 1.95  $&0.39$\pm$0.06&$ 37.09\pm0.06^\star $&&\\
GH163            & VLA-A   &$1.4$&$ 739  $&$ 620    $&0.92$\pm$0.11&$ 37.58\pm0.03^\diamond $&$-0.99\pm0.04^{\ddagger}$&$0.67\pm0.11$\\
GH163            & VLA-A   &$9$&$   190  $&$ 102    $&0.91$\pm$0.04&$ 37.46\pm0.06^\bullet $&$-0.55\pm0.27^{\dagger}$&\\
\enddata
\tablecomments{Columns give (1) source alias, (2) telescope, (3) frequency, (4-5) major and minor axis of the beam, in physical scale (parsec), (6) the concentration index, (7) C-band (5\,GHz) luminosity, (8) spectral index and (9) the fraction of diffuse and extended radio emission between the compact VLA 1.4\,GHz emission and the total VLBA 1.5\,GHz emission over the compact VLA 1.4\,GHz radio emission, defined as $\frac{S_{p,\mathrm{VLA,1.4GHz}}-S_{i,\mathrm{VLBA,1.5GHz}}}{S_{p,\mathrm{VLA,1.4GHz}}}$, here we assume the VLA 1.4\,GHz and VLBA 1.5\,GHz are at approximately equally frequency band.}
\tablecomments{C-band Radio luminosity. $\star$: estimated by taking the VLBA L-band radio flux density and the VLA A-array L/X spectral index; $\diamond$: estimated by taking the VLA A-array L-band radio flux density and the VLA A-array L/X spectral index; $\bullet$: estimated by taking the VLA A-array X-band radio flux density and the in-band VLA A-array wide X-band spectral index.}
\tablecomments{Radio spectral index. $\ddagger$: measured between VLA A-array L-band and X-band; $\dagger$: the in-band spectral index obtained from the wide X-band spans from 8.5 to 9.5\,GHz.}
\end{deluxetable*}

\begin{deluxetable*}{ccccccc}
\tablecaption{Emission properties of the large scale region.\label{tab:largescaleemission}}
\tablewidth{0pt}
\tablehead{
\colhead{Source} & \colhead{$z$} & \colhead{$F_\nu$} & \colhead{$p$} & \colhead{$R$} & \colhead{$B$} & \colhead{$E$}\\
\colhead{} & \colhead{} & \colhead{(mJy)} & \colhead{} & \colhead{(pc)} & \colhead{(mG)} & \colhead{($\times 10^{53}$ erg)} }
\startdata
GH047 & 0.025 & 0.104 & 2.3 & 33.8 & 0.13 & 0.05 \\
GH106 & 0.033 & 2.425 & 3.0 & 350.2 & 0.06 & 12.14 \\
GH158 & 0.045 & 0.065 & 2.6 & 50.4 & 0.12 & 1.26 \\
GH163 & 0.021 & 0.215 & 3.0 & 47.7 & 0.12 & 1.15 \\
\hline
\enddata
\tablecomments{The columns are (1) source alias, (2) source redshift, taken from Table \ref{tab:info}, (3) total flux density of the emitting region $F_\nu = S_i - S_p$ based on the reported VLA L-band flux densities in Table \ref{tab:results}, (4) electron energy index $p = 1-2 \alpha$ where the spectral indices $\alpha$ are the estimates from the L/X band data in Table \ref{tab:results}, (5) emitting region size evaluated as $R = R_{min,{\rm VLA}} (1-S_p/S_i)-R_{min,{\rm VLBA}}$, (6) magnetic field strength in the emitting region $B$, and (7) total energy in the emitting region, $E$.}
\end{deluxetable*}

\end{document}